\documentclass{PoS}
\usepackage{amsmath}

\newcommand{\be}{\begin{eqnarray}}
\newcommand{\ee}{\end{eqnarray}}
\newcommand{\ba}{\begin{array}}
\newcommand{\ea}{\end{array}}

\title{Mechanical properties of  particles}

\ShortTitle{Mechanical properties of  particles }

\author{\speaker{Maxim V. Polyakov}
\\
  Ruhr University Bochum, Faculty of Physics and Astronomy,
Institute of Theoretical Physics II, D-44780 Bochum, Germany,\\
National Research Centre ``Kurchatov Institute'': Petersburg Nuclear Physics
Institute,
		 RU-188300 Gatchina, Russian Federation     \\
        E-mail: \email{maxim.polyakov@rub.de}}

\author{{Peter Schweitzer}
\\
 Department of Physics, University of Connecticut, 
		Storrs, CT 06269, USA    \\
        E-mail: \email{peter.schweitzer@phys.uconn.edu }}

\abstract{ Selected topics related to the physics of the energy-momentum tensor (EMT) form factors  are discussed. The topics are: 1) Fundamental mechanical properties of particles and gravity
2) Mechanical properties of non-spherical particles  3) Gravitational form factors of Goldstone bosons  4) Nucleon's seismology?}

\FullConference{23rd International Spin Physics Symposium - SPIN2018 -\\
		10-14 September, 2018\\
		Ferrara, Italy}

\begin{document}
\section*{\normalsize \bf Introduction}

The gravitational form factors are related to very wide class of physics problems ranging from the fundamental questions of General Relativity to the theory of hard exclusive processes
and physics of exotic charmonia. Although the direct access to these form factors with gravitational forces (at least  with {those} available in the Solar system) is out of reach, their first measurements
in hard QCD processes became available \cite{Kumericki:2015lhb,Kumano:2017lhr,Nature}.  In recent paper~\cite{Polyakov:2018zvc} a detailed review of the theory and the phenomenology
of the gravitational form factors is provided, {and a} comprehensive list of related literature is given.

The field is fastly developing, during last few months several important results were obtained:
\begin{itemize}
\item
In Ref.~\cite{Lorce:2018egm} the definition of the force distributions inside hadrons in wide range of reference frames is provided. Interesting
connections between gravitational form factors and physics of compact stars are discussed.

\item
New lattice measurements of the pressure and shear force distributions (for quarks and gluons) in the pion and in the nucleon were reported in Refs.~\cite{Shanahan:2018nnv,Shanahan:2018pib}.
We note however that the results for the shear force distribution $s(r)$ obtained in these papers violate the stability constraint $s(r)>0$. This needs a clarification.

\item
The comprehensive perturbative QCD analysis of the trace anomaly for {\it quark} EMT was performed in Refs.~\cite{Hatta:2018sqd,Tanaka:2018nae}.
This analysis is very important for the discussion of the nucleon mass decomposition in QCD.

\end{itemize}
In this short contribution we touch only several points (sometimes speculative) which are covered neither in the review \cite{Polyakov:2018zvc} nor in recent literature. 

\section*{\normalsize \bf Fundamental mechanical properties of particles and gravity}

Our intuitive perception of the mass is related to the gravity (weighing experiment). The gravity itself is equivalent to a non-trivial space-time metric $g_{\mu\nu}(x)$
(we shall consider here the Minkowski metric signature $\eta_{\mu\nu}={\rm diag}(+---)$).  The mass resulting from  the weighing experiment is related to the variation of the action in respect to the
static $g_{00}(\bf r)$. The corresponding expression has the form:
\be
\label{eq:mass}
M=\int d^3 {\bf r} \frac{2}{\sqrt{-g}} \frac{\delta S}{\delta g_{00}({\bf r})}\Big|_{g_{\mu\nu}=\eta_{\mu\nu}} .
\ee
Another basic mechanical property, the total angular momentum $J^i$ (particle spin), is obtained by the variation of the action in respect to static $g_{0i}({\bf r})$:
\be
\label{eq:spin}
J^i=\epsilon^{ikl} \int d^3 {\bf r}\  r^k \frac{2}{\sqrt{-g}} \frac{\delta S}{\delta g_{0l}({\bf r})}\Big|_{g_{\mu\nu}=\eta_{\mu\nu}} .
\ee  
The classical application of this equation is the measurement of the Earth rotation by the Foucault pendulum. 

The mass and the angular momentum (spin) of a particle are well known and frequently discussed fundamental characteristics of  particles. There is another -- not frequently discussed --
fundamental characteristic of a particle, which is related to the variation of the action in respect to  the spatial metric $g_{ik}({\bf r})$. Such variation corresponds
to the change of the 3D distances, i.e. to the deformation of the particle and hence to its elasticity properties. Therefore, we can introduce the quantity 
 \cite{Polyakov:2002yz}:
\be
\label{eq:Dtermdef}
D= -\frac{2 M}{5} \int d^3 {\bf r} \left( r^i r^k-\frac 13 r^2 \delta^{ik}\right) \frac{2}{\sqrt{-g}} \frac{\delta S}{\delta g_{ik}({\bf r})}\Big|_{g_{\mu\nu}=\eta_{\mu\nu}} ,
\ee
called $D$-term \footnote{The name ``$D$-term" is rather technical, it can be traced back to more or less accidental
 notations chosen in Ref.~\cite{Polyakov:1999gs}. Nowadays, given more clear physics meaning of this quantity, we might call this term as ``{\it Druck-term}" derived from 
German word for ``pressure".  }, it characterises the distribution of internal forces in the particle. 
The $D$-term is an intrinsic characteristic of any particle, which is as fundamental as particle mass and spin. 
The detailed discussion of the $D$-term for various systems can be found in recent review \cite{Polyakov:2018zvc}. The stability of the system requires
that the $D$-term is negative \cite{Polyakov:2018zvc}, and indeed in all known examples $D<0$, even for the unstable particles. 

The effective action for the nucleon (described by the spinor field $N(x)$) interacting with the external gravitational field can be written as:
\be
\label{eq:Seff}
S_{\rm eff}=\int d^4 x \ \sqrt{-g} \left( \overline{N}(x)(i \gamma^\mu {\cal D}_\mu- M)N(x)  - \frac{D}{8 M}\ R(x)\ \overline N(x) N(x) + \ldots \right),
\ee  
where $R(x)$ is the scalar curvature of the space-time, ${\cal D}_\mu=\partial_\mu+\frac18 [\gamma_A,\gamma_B] \omega^{AB}_\mu$ is the covariant derivative written in terms of vierbeins and spin-connection
(see e.g. \cite{Diakonov:2011fs}), and ellipsis 
stays for higher order terms\footnote{These terms can contain contributions  of the type $R^{\mu\nu} {\cal D}_\mu \overline{N} {\cal D}_\nu{N}$, etc. The detailed classification of all
terms in the effective action will be given elsewhere. }. Using this effective action and eqs. (\ref{eq:mass},\ref{eq:spin},\ref{eq:Dtermdef}) one obtains that the nucleon mass $M_N=M$, its spin $J^i_N=\frac{\sigma^i}{2}$
and the nucleon $D$-term $D_N=D$.

From expression for the effective action (\ref{eq:Seff}) we see that the $D$-term enters multiplied by the scalar curvature of the space-time, so it is very strongly suppressed 
in gravitational fields available in Solar system. However, this term might play essential role in physics of hadrons in recently observed violent events such as the neutron stars mergers
\cite{TheLIGOScientific:2017qsa}.

\noindent
\section*{\normalsize \bf Mechanical properties of non-spherical particles}

Particles with spins $J=0,\frac 12$ posses the spherical symmetry.  The spherical symmetry allows to express the static stress tensor:
\be
\Theta^{ik}({\bf r})=\frac{2}{\sqrt{-g}} \frac{\delta S}{\delta g_{ik}({\bf r})},
\ee
in terms of pressure ($p(r)$) and shear forces ($s(r)$) distribution inside the particle:
\be
\label{eq:emtstatic}
\Theta^{ik}({\bf r})=p(r)\ \delta^{ik}+s(r)\ Y_2^{ik}.
\ee 
Here we introduce the irreducible (symmetric and traceless) tensor  of $n$-th rank:
\be
Y_{n}^{i_1 i_2 ... i_n} = \frac{(-1)^n}{(2 n-1)!!} r^{n+1} \partial^{i_1}...\partial^{i_n} \frac{1}{r}, 
\quad \mbox{i.e.} \quad
Y_0=1,\ Y_1^{i}=\frac{r^{i}}{r}, \ Y_2^{ik}=\frac{r^{i} r^{k}}{r^2}-\frac13 \delta^{ik}, \ {\rm etc.}
\ee 
The pressure and shear forces in eq.~(\ref{eq:emtstatic}) can be related to the Fourier transform of the EMT form factor in the Breit frame
 ($\widetilde{D}(r)=\int \frac{d^3 {\Delta}}{(2\pi)^3}e^{-i { \Delta\cdot r}} D(-\Delta^2)$) \cite{Polyakov:2002yz,Polyakov:2018zvc}\footnote{See recent
paper \cite{Lorce:2018egm} for detailed discussion of the definition of pressure and shear forces in wide class of reference frames. }
\be
p(r)=\frac{1}{6M} \frac{1}{r^2} \frac{d}{dr}r^2\frac{d}{dr} \widetilde{D}(r)=\frac{1}{6M} \partial^2 \   \widetilde{D}(r),\quad s(r)=-\frac{1}{4M} r \frac{d}{dr}\frac{1}{r}\frac{d}{dr} \widetilde{D}(r).
\ee
These relations follow from the equilibrium equation $\partial_k \Theta^{ik}({\bf r})=0$ and they {guarantee} the general stability condition \cite{LLv7}:
\be
\label{eq:stabilitygeneral}
\int d^3 {\bf r}\  \Theta^{ik}({\bf r}) =0,
\ee
due to obvious relations:
\be
\int d^3 {\bf r}\  p(r) =\frac{1}{6M}\int d^3 {\bf r}\ \partial^2\   \widetilde{D}(r) =0, \quad \int d\Omega\ Y_2^{ik}=0.
\ee

For higher spin particles $J\ge 1$ more terms in expression for the static stress tensor $\Theta^{ik}({\bf r})$ are allowed. The new terms can be classified in
terms of multipole expansion. General expansion to the quadrupole order has the form:
\be
\label{eq:quadrupole}
\Theta^{ik}({\bf r}) =p_0(r) \delta^{ik}+s_0(r)Y_2^{ik} + p_2(r) \hat{Q}^{ik}+2 s_2(r) 
\left[\hat{Q}^{ip}Y_{2}^{pk}+\hat Q^{kp}Y_{2}^{pi} -\delta^{ik} \hat Q^{pq}Y_{2}^{pq}    \right]+\ldots
\ee
Here ellipsis stays for the contribution of $2^n$- multipoles with $n>2$ parametrised by $p_n(r), s_n(r)$, etc. The quadrupole operator is {the} $(2J+1)\times (2J+1)$
matrix:
\be
\hat Q^{ik}=\frac 12 \left( \hat J^{\ i} \hat J^{\ k}+\hat J^{\ k} \hat J^{\ i} -\frac 23 J(J+1) \delta^{ik} \right),
\ee
which is expressed in terms of the spin operator $\hat J^{\ i}$. The spin operator can be expressed in terms of the SU(2) Clebsch-Gordan coefficients (in the spherical basis):
\be
\hat J^{\ \mu}_{m'm}= \sqrt{J(J+1)}\  C_{J m 1\mu}^{J m'}.
\ee
The quadrupole pressure and shear forces distributions ($p_2(r),s_2(r)$) can be expressed through the Fourier transform of additional
EMT form factors  for higher spin particles\footnote{Generically, for an integer spin $J$ particle,  the EMT has $4 J+2$ form factors, plus  $2 J+1$ additional form factors for individual quark and gluon EMTs \cite{PSS} .
The {latter} describe the non-conservation of individual EMTs.} :
\be
p_2(r)=\frac{1}{6M} \frac{1}{r^2} \frac{d}{dr}r^2\frac{d}{dr} \widetilde{E}(r)=\frac{1}{6M} \partial^2\   \widetilde{E}(r),\quad s_2(r)=-\frac{1}{4M} r \frac{d}{dr}\frac{1}{r}\frac{d}{dr} \widetilde{E}(r).
\ee
This form of quadrupole pressure and shear forces is the consequence of the equilibrium equation $\partial_k \Theta^{ik}({\bf r})=0$ which guaranties the 
stability condition (\ref{eq:stabilitygeneral}).

For a particle of arbitrary spin we can introduce more general tensor quantities:
\be
\label{eq:massmulti}
M_n^{k_1\ldots k_n}=\int d^3 {\bf r}\   r^{n}\ Y_n^{k_1 \ldots k_n} \ \Theta^{00}({\bf r}),
\ee   
which correspond {to} $2^n$-multipoles of the energy distribution, obviously  $M_0=M$. Note, that only even $n$ are allowed by the $P$-parity conservation.
Eq.~(\ref{eq:massmulti}) can be reformulated as the multipole expansion of the energy density :
\be
\Theta^{00}({\bf r}) =\sum_{n=0,2,\ldots} \epsilon_n(r) \hat Q_n^{k_1\ldots k_n}\ Y_n^{k_1\ldots k_n},
\ee
where $\hat Q_n^{k_1\ldots k_n}$ is the  $2^n$-pole spin operator and $\epsilon_n(r)$ is the corresponding $2^n$-pole energy density.

Analogously, for an arbitrary spin particle we can introduce a set of dimensionless tensors of rank $n+2$:
\be
\label{eq:dtensor}
D_{n}^{ik k_1 k_2 ...k_n}= -\frac{4}{M} \int d^3 {\bf r}\   (M r)^{n} Y_n^{k_1 k_2 ...k_n}  \ \Theta^{ik}({\bf r}).
\ee
Again, only even $n$ are allowed by the $P$-parity and $D_0^{ik}=0$ due to the stability condition (\ref{eq:stabilitygeneral}). For particles with spin $J=0,\frac 12$
only $D_2^{ikk_1k_2}$ is non-zero and can be expressed through the $D$-term (\ref{eq:Dtermdef}):
\be
D_2^{ikk_1k_2}=\left( \delta^{ik_1}\delta^{kk_2}+\delta^{kk_1}\delta^{ik_2}-\frac 23 \delta^{ik}\delta^{k_1k_2} \right)\ D.
\ee
The tensor observables (\ref{eq:dtensor}) can be related to GPDs, see e.g. the discussion for spin-1 hadrons in recent paper~\cite{Cosyn:2018thq}.

\section*{Gravitational form factors of Goldstone bosons}

Goldstone bosons of a spontaneously broken symmetry  in any theory play crucial role in dynamics of the theory. 
For example, the phenomenon of spontaneous breakdown of the chiral symmetry in the strong interaction is crucial for 
the description of the mass spectrum and dynamics in QCD.

The Goldstone bosons of spontaneously broken chiral symmetry are (almost) massless spin-0 particles and therefore 
the $D$-term cannot be defined in terms of static stress tensor, see (\ref{eq:Dtermdef}). 
For {Goldstone bosons we define the $D$-term} in Lorentz covariant way, in terms of 
EMT form factors:
\begin{equation}
     \langle p^\prime|  \Theta^{\mu\nu}_a(0) |p\rangle
    = 
      2 \,P^\mu P^\nu A^a(t)
    + \frac{1}{2} \left({\Delta^\mu\Delta^\nu-\eta^{\mu\nu}\Delta^2} \right) D^a(t)
    +2 f_\pi^2 \eta^{\mu\nu} \,{\bar c}^a(t)    \label{eq:EMT-FFs-spin-1/2}
\end{equation}
Here $P=(p'+p)/2$, $\Delta = p'-p$ and $f_\pi$ is the 
pion decay constant which {has dimension of mass and} sets the mass scale in the effective theory.  
We introduced the form factors for individual quark and gluon EMTs.
The total EMT is conserved
\be\label{Eq:EMT-cons}
	\partial_\mu \Theta^{\mu\nu} = 0, \quad \quad
	 \Theta^{\mu\nu} = \sum_q \Theta^{\mu\nu}_q+ \Theta^{\mu\nu}_g \; ,
\ee 
hence $\sum_{a={q,g}}\bar{c}^a(t)=0$.
The quark form factor ${\bar c}^Q(t)=\sum_{a=u,d,s, \ldots}\bar{c}^a(t)=-\bar c^g(t)$,
 describes the non-conservation of EMT for individual quark and gluon pieces. 
 This form factor is important to determine the pressure forces distribution in a hadron individually for
quarks and gluons, and to study the forces between quark and gluon subsystems in hadrons\footnote{\noindent 
The stability equation for the quark part of the stress tensor has the form:
\be
\nonumber
\frac{\partial \Theta^{ik}_Q({\bf r})}{\partial r^k} +f^i({\bf r})=0. 
\ee
This equation can be interpreted (see e.g \S 2 of \cite{LLv7}) as the equilibrium 
equation for quark internal stress and external force (per unit of the volume) $f^i({\bf r})$ acting on quark subsystem from the side of the gluons.
 This gluon force can be 
computed in terms of EMT form factor $\bar c^Q(t)$ as \cite{Polyakov:2018exb}:
\be
\nonumber
f^i({\bf r})=M  \frac{\partial}{\partial r^i} \int {\frac{d^3\Delta}{(2\pi)^3}}\ e^{{-i} {\Delta \cdot r}}\
	 \bar c^Q(-{\Delta}^2).
\ee
The total squeezing  (stretching) gluon force acting on quarks in the nucleon is equal to \cite{Polyakov:2018exb}:
\be
\nonumber
F_{\rm total}= \frac{2 M}{ \pi} \int_{-\infty}^0 \frac{dt}{\sqrt{-t}}\  \bar c^Q(t).
\ee
Estimates in the instanton model of the QCD vacuum in Ref.~\cite{Polyakov:2018exb} show that this force is squeezing and have rather small size
$F_{\rm total}\simeq {6}\cdot 10^{-2}$~GeV/fm.
} 
(see recent discussions in
 \cite{Tanaka:2018wea,Polyakov:2018exb,Hatta:2018sqd,Tanaka:2018nae}).

The form factors in eq.~(\ref{eq:EMT-FFs-spin-1/2}) at zero momentum transfer can be fixed by the soft pion theorem:
\be
\lim_{p^{\prime \mu} \to 0} \langle p^\prime|  \Theta^{\mu\nu}_Q(x) |p\rangle =0.
\ee 
This theorem leads to the relation among form factors:
\be
0= \frac12 p^\mu p^\nu  \left(A^Q(0) +D^Q(0) \right) +2 f_\pi^2\  \eta^{\mu\nu} \bar c^Q(0).
\ee
This equation is satisfied if the EMT form factors of massless Goldstone boson are related to each other by:
\be
D^Q(0)=-A^Q(0), \quad \bar c^Q(0)=0.
\ee
From the first equality we obtain immediately the value of the $D$-term of the pion in the chiral limit $D=-1$ \cite{Voloshin:1982eb}.

Our result that $\bar c^Q(0)=0$ for Goldstone bosons is valid for arbitrary QCD normalisation point. Therefore, it is in contradiction with the result of Ref.~\cite{Hatta:2018sqd}
for this form factor at asymptotically large QCD normalisation point:
\be
\lim_{\mu\to\infty} \bar c^Q(0)=\frac 14 \left(\frac{N_f}{4 C_F+N_f}+\frac{2 N_f}{\beta_0} \right) \neq 0.
\ee
It is important to find a reason for this contradiction.

\section*{\boldmath Nucleon's seismology?}

{Up to now we consider the energy density $\Theta^{00}({\bf r})$ and distribution of forces encoded in the stress tensor $\Theta^{ik}({\bf r})$ separately.
It would be interesting to establish connection between these quantities, this would be a step towards an understanding of the equation
of state inside a hadron. 
If we treat the interior of a hadron as an elastic medium and boldly identify elastic moduli $K$ and $\mu$ (see \S 4 of \cite{LLv7}) with the pressure and shear forces
distributions as $K=p(r)$ and $2\mu=s(r)$, we can obtain
the longitudinal ($c_l$ ) and transverse ($c_t$) speeds of elastic wave\footnote{The seismic waves are well known examples of this phenomenon.} (see \S 22 of \cite{LLv7}):
\be
\label{eq:speedof sounds}
c_l(r) =\sqrt{\frac{\frac 23 s(r)+p(r)}{\Theta^{00}({\bf r})}}, \quad c_t(r) =\sqrt{\frac{s(r)}{2 \Theta^{00}({\bf r})}}.
\ee
These relations demonstrate once again that $\frac 23 s(r)+p(r)>0$  and $s(r)>0$, which corresponds to the stability conditions.
Using the chiral perturbation theory one obtains for the nucleon that at 
large distances $r\to\infty$:
\be
c_l(r)\to \sqrt{\frac 13}, \quad c_t(r)\to \sqrt{\frac 12},
\ee
in the chiral limit, and
\be
c_l(r)\sim \sqrt{\frac{1}{m_\pi r}}\to 0, \quad c_t(r)\to \sqrt{\frac 12},
\ee
for $m_\pi\neq 0$. 
Imposing the conditions that the speeds of elastic waves are less than the speed of light we obtain the following inequalities:
\be
\label{eq:ineqT00sp}
\Theta^{00}({\bf r})-\left[\frac 23 s(r)+p(r)\right]>0, \quad \Theta^{00}({\bf r})-\frac 12 s(r)>0 .
\ee
From these inequalities we can obtain the low bound  (upper bound for the absolute value) for the allowed value of the $D$-term:
\be
0\geq D\geq -\frac{8}{15} M^2 \langle r^2\rangle_{E} ,\ {\rm or }\  \  |D|\leq \frac{8}{15} M^2 \langle r^2\rangle_{E} ,
\ee 
where $\langle r^2\rangle_{E}$ is the mean square radius of the energy density defined by:
\be
\langle r^2\rangle_{E}=\frac 1M \int d^3{\bf  r} \ r^2\ \Theta^{00}({\bf r})
\ee
It is very interesting that the inequalities (\ref{eq:ineqT00sp}) are 
satisfied in various models. 
For example, in the Skyrme model we have {for radially symmetric solutions}
\begin{eqnarray}
&&\Theta^{00}({\bf r})-\left[\frac 23 s(r)+p(r)\right] =\frac{\sin^2(F(r))}{r^2}\left(\frac{f_\pi^2}{2}+\frac{\sin^2(F(r))}{e^2 r^2}\right)+\frac{f_\pi^2 m_\pi^2}{2}\left(1-\cos(F(r))\right)\geq 0,\\
&&\Theta^{00}({\bf r})-\frac 12 s(r)=\frac{\sin^2(F(r))}{r^2}\left(\frac{3f_\pi^2}{8}+\frac{\sin^2(F(r))}{e^2 r^2}+\frac{1}{2 e^2} F'(r)^2\right)+\frac{f_\pi^2 m_\pi^2}{2}\left(1-\cos(F(r))\right)\geq 0,
\nonumber
\end{eqnarray}
{where the profile function $F(r)$ satisfies $F(0)=B\,\pi$ ($B=\,$winding 
number) and vanishes for $r\to\infty$.}
Both expressions are explicitly {\it positive} and hence the general inequalities (\ref{eq:ineqT00sp}) are satisfied automatically. 
Numerical studies of Skyrmions, show that for winding number $B=1$ Skyrmion $c_l\leq \sqrt{1/3}$
and $c_t\leq \sqrt{1/2}$ whereas for {radially symmetric} Skyrmions with higher winding 
numbers both velocities reach from below speed of light at $(B-1)$ points inside the Skyrmion.  
Note  that {radially symmetric} Skyrmions with $B\ge 2$ are unstable. Therefore,  we come to the conjecture that the inequalities:
\be
\label{eq:conj}
c_l\leq \sqrt{\frac 13}, \quad
c_t\leq \sqrt{\frac 12},
\ee
might be considered as the criteria for the stability of the light baryons and the nuclei\footnote{We note that this conjecture being 
applied to the liquid drop gives for the equation of state $p\leq 1/3  \epsilon$, where $p$
is the pressure and $\epsilon$ is the energy density in the drop}. 
Recently similar inequalities were discussed in Ref.~\cite{Lorce:2018egm}. In that paper they were
related to the energy conditions which reflect the principles of relativity and play an important role in General Relativity.
Would be interesting to find the connection with consideration here. 

Our conjecture is still very speculative, but if it is true, it leads to the following bound
on the absolute value of the $D$-term:
 \be
 \label{eq:Dbound}
 |D|\leq \frac{2}{9}\  M^2 \langle r^2\rangle_{E} .
 \ee

{Numerical studies of the Q-balls (see discussion of EMT for Q-balls in \cite{Mai:2012yc,Mai:2012cx}) shows that the inequalities (\ref{eq:ineqT00sp}) 
as well as the bound (\ref{eq:Dbound}) are always satisfied, but the conjectured  inequalities
(\ref{eq:conj}) are violated even for the stable Q-balls. This violation happens only in the small region close to the centre of the soliton. 
It would be important to identify the class of systems for which our conjecture (\ref{eq:conj}) is valid. At least for the nucleon as
a Skyrmion it is  true.
}

\noindent
\section*{\normalsize \bf Acknowledgements}

\noindent
The results presented here would be impossible without  discussions and close collaboration with Kirill Semenov-Tian-Shansky, Hyeon-Dong Son, and Bao-Dong Sun. 
We are also  grateful to C\'edric Lorc\'e  and Oleg Teryaev for illuminating discussions.
This work is supported by the Sino-German CRC 110
``Symmetries and the Emergence of Structure in QCD".



\begin{thebibliography}{99}


 \bibitem{Kumericki:2015lhb}
  K.~Kumericki and D.~Mueller,
  ``Description and interpretation of DVCS measurements,''
  EPJ Web Conf.\  {\bf 112} (2016) 01012
  [arXiv:1512.09014 [hep-ph]].

\bibitem{Kumano:2017lhr}
  S.~Kumano, Q.~T.~Song and O.~V.~Teryaev,
  ``Hadron tomography by generalized distribution amplitudes in pion-pair production process $\gamma^* \gamma \rightarrow \pi^0 \pi^0 $ and gravitational form factors for pion,''
  Phys.\ Rev.\ D {\bf 97} (2018) no.1,  014020
  [arXiv:1711.08088 [hep-ph]].

  
  \bibitem{Nature}
   V.~D.~Burkert, L.~Elouadrhiri and F.~X.~Girod,
  ``The pressure distribution inside the proton,''
  Nature {\bf 557} (2018) no.7705,  396.



\bibitem{Polyakov:2018zvc}
  M.~V.~Polyakov and P.~Schweitzer,
  ``Forces inside hadrons: pressure, surface tension, mechanical radius, and all that,''
  Int.\ J.\ Mod.\ Phys.\ A {\bf 33} (2018) no.26,  1830025
  [arXiv:1805.06596 [hep-ph]].


  \bibitem{Lorce:2018egm}
  C.~Lorce, H.~Moutarde and A.~P.~Trawinski,
  ``Revisiting the mechanical properties of the nucleon,''
  arXiv:1810.09837 [hep-ph].

\bibitem{Shanahan:2018nnv}
  P.~E.~Shanahan and W.~Detmold,
  ``The pressure distribution and shear forces inside the proton,''
  arXiv:1810.07589 [nucl-th].

\bibitem{Shanahan:2018pib}
  P.~E.~Shanahan and W.~Detmold,
  ``Gluon gravitational form factors of the nucleon and the pion from lattice QCD,''
  arXiv:1810.04626 [hep-lat].

\bibitem{Hatta:2018sqd}
  Y.~Hatta, A.~Rajan and K.~Tanaka,
  ``Quark and gluon contributions to the QCD trace anomaly,''
  arXiv:1810.05116 [hep-ph].

\bibitem{Tanaka:2018nae}
  K.~Tanaka,
  ``Three-loop formula for quark and gluon contributions to the QCD trace anomaly,''
  arXiv:1811.07879 [hep-ph].


\bibitem{Polyakov:2002yz}
  M.~V.~Polyakov,
  ``Generalized parton distributions and strong forces inside nucleons and nuclei,''
  Phys.\ Lett.\ B {\bf 555} (2003) 57
  [hep-ph/0210165].
  
  \bibitem{Polyakov:1999gs}
  M.~V.~Polyakov and C.~Weiss,
  ``Skewed and double distributions in pion and nucleon,''
  Phys.\ Rev.\ D {\bf 60} (1999) 114017
  [hep-ph/9902451].
  
  
  
   
  
  \bibitem{Diakonov:2011fs}
  D.~Diakonov, A.~G.~Tumanov and A.~A.~Vladimirov,
  ``Low-energy General Relativity with torsion: A Systematic derivative expansion,''
  Phys.\ Rev.\ D {\bf 84} (2011) 124042
  [arXiv:1104.2432 [hep-th]].
  
  \bibitem{TheLIGOScientific:2017qsa}
  B.~P.~Abbott {\it et al.} [LIGO Scientific and Virgo Collaborations],
  ``GW170817: Observation of Gravitational Waves from a Binary Neutron Star Inspiral,''
  Phys.\ Rev.\ Lett.\  {\bf 119} (2017) no.16,  161101
  [arXiv:1710.05832 [gr-qc]].
  
 
   \bibitem{PSS}
M.V.~Polyakov, K.M.~Semenov-Tian-Shansky, Bao-Dong Sun, ``Gravitational form factors for particles with arbitrary spin"", in preparation.

  
  \bibitem{Cosyn:2018thq}
  W.~Cosyn, A.~Freese and B.~Pire,
  ``Polynomiality sum rules for generalized parton distributions of spin-1 targets,''
  arXiv:1812.01511 [hep-ph].

  
  \bibitem{LLv7}
  L.~D.~Landau and E.~M.~Lifshitz,
  Course of Theoretical Physics, vol. VII  ``Theory of Elasticity,''
  %
  Pergamon Press, 1970.

  
\bibitem{Tanaka:2018wea}
  K.~Tanaka,
  ``Operator relations for gravitational form factors of a spin-0 hadron,''
  Phys.\ Rev.\ D {\bf 98} (2018) no.3,  034009
  [arXiv:1806.10591 [hep-ph]].
  
\bibitem{Polyakov:2018exb}
  M.~V.~Polyakov and H.~D.~Son,
  ``Nucleon gravitational form factors from instantons: forces between quark and gluon subsystems,''
  JHEP {\bf 1809} (2018) 156
   [JHEP {\bf 2018} (2020) 156]
  [arXiv:1808.00155 [hep-ph]].
  
   
  \bibitem{Voloshin:1982eb}
  M.~B.~Voloshin and A.~D.~Dolgov,
  ``On Gravitational Interaction Of The Goldstone Bosons,''
  Sov.\ J.\ Nucl.\ Phys.\  {\bf 35} (1982) 120
   [Yad.\ Fiz.\  {\bf 35} (1982) 213].
  
 
\bibitem{Mai:2012yc}
  M.~Mai and P.~Schweitzer,
  ``Energy momentum tensor, stability, and the D-term of Q-balls,''
  Phys.\ Rev.\ D {\bf 86} (2012) 076001
  [arXiv:1206.2632 [hep-ph]].\\
\bibitem{Mai:2012cx}
M.~Mai and P.~Schweitzer,
  ``Radial excitations of Q-balls, and their D-term,''
  Phys.\ Rev.\ D {\bf 86} (2012) 096002
  [arXiv:1206.2930 [hep-ph]].





 \end{thebibliography}
\end{document}